\documentstyle[epsfig,12pt,a4]{article}
\newcommand{\pipi}{\mbox{$\pi^+\pi^-$ }}
\newcommand{\pipizer}{\mbox{$\pi^0\pi^0$ }}

\begin{document}
\begin{titlepage}
\def\footnoterule{\hrule width 1.0\columnwidth}
\begin{tabbing}
put this on the right hand corner using tabbing so it looks
 and neat and in \= \kill
\> {10 February 1999}
\end{tabbing}
\bigskip
\bigskip
\begin{center}{\Large  {\bf A partial wave analysis of the centrally produced
$\pi^0 \pi^0$ system
in pp interactions at 450 GeV/c}
}\end{center}
\bigskip
\bigskip
\begin{center}{        The WA102 Collaboration
}\end{center}\bigskip
\begin{center}{
D.\thinspace Barberis$^{  4}$,
W.\thinspace Beusch$^{   4}$,
F.G.\thinspace Binon$^{   6}$,
A.M.\thinspace Blick$^{   5}$,
F.E.\thinspace Close$^{  3,4}$,
K.M.\thinspace Danielsen$^{ 10}$,
A.V.\thinspace Dolgopolov$^{  5}$,
S.V.\thinspace Donskov$^{  5}$,
B.C.\thinspace Earl$^{  3}$,
D.\thinspace Evans$^{  3}$,
B.R.\thinspace French$^{  4}$,
T.\thinspace Hino$^{ 11}$,
S.\thinspace Inaba$^{   8}$,
A.V.\thinspace Inyakin$^{  5}$,
T.\thinspace Ishida$^{   8}$,
A.\thinspace Jacholkowski$^{   4}$,
T.\thinspace Jacobsen$^{  10}$,
G.T\thinspace Jones$^{  3}$,
G.V.\thinspace Khaustov$^{  5}$,
T.\thinspace Kinashi$^{  12}$,
J.B.\thinspace Kinson$^{   3}$,
A.\thinspace Kirk$^{   3}$,
W.\thinspace Klempt$^{  4}$,
V.\thinspace Kolosov$^{  5}$,
A.A.\thinspace Kondashov$^{  5}$,
A.A.\thinspace Lednev$^{  5}$,
V.\thinspace Lenti$^{  4}$,
S.\thinspace Maljukov$^{   7}$,
P.\thinspace Martinengo$^{   4}$,
I.\thinspace Minashvili$^{   7}$,
T.\thinspace Nakagawa$^{  11}$,
K.L.\thinspace Norman$^{   3}$,
J.P.\thinspace Peigneux$^{  1}$,
S.A.\thinspace Polovnikov$^{  5}$,
V.A.\thinspace Polyakov$^{  5}$,
V.\thinspace Romanovsky$^{   7}$,
H.\thinspace Rotscheidt$^{   4}$,
V.\thinspace Rumyantsev$^{   7}$,
N.\thinspace Russakovich$^{   7}$,
V.D.\thinspace Samoylenko$^{  5}$,
A.\thinspace Semenov$^{   7}$,
M.\thinspace Sen\'{e}$^{   4}$,
R.\thinspace Sen\'{e}$^{   4}$,
P.M.\thinspace Shagin$^{  5}$,
H.\thinspace Shimizu$^{ 12}$,
A.V.\thinspace Singovsky$^{ 1,5}$,
A.\thinspace Sobol$^{   5}$,
A.\thinspace Solovjev$^{   7}$,
M.\thinspace Stassinaki$^{   2}$,
J.P.\thinspace Stroot$^{  6}$,
V.P.\thinspace Sugonyaev$^{  5}$,
K.\thinspace Takamatsu$^{ 9}$,
G.\thinspace Tchlatchidze$^{   7}$,
T.\thinspace Tsuru$^{   8}$,
M.\thinspace Venables$^{  3}$,
O.\thinspace Villalobos Baillie$^{   3}$,
M.F.\thinspace Votruba$^{   3}$,
Y.\thinspace Yasu$^{   8}$.
}\end{center}

\begin{center}{\bf {{\bf Abstract}}}\end{center}

{
A partial wave analysis of the centrally produced \pipizer
channel has been performed in $pp$ collisions using
an incident beam momentum of 450~GeV/c.
An unambiguous physical solution has been found.
Evidence is found for the $f_0(980)$, $f_0(1300)$ and $f_0(1500)$ in the
the S-wave.
and the $f_2(1270)$ is observed dominantly in the $D_0^-$-wave.
In addition, there is evidence for a
broad enhancement in the D-wave below 1 GeV.
}
\bigskip
\bigskip
\bigskip
\bigskip\begin{center}{{Submitted to Physics Letters}}
\end{center}
\bigskip
\bigskip
\begin{tabbing}
aba \=   \kill
$^1$ \> \small
LAPP-IN2P3, Annecy, France. \\
$^2$ \> \small
Athens University, Physics Department, Athens, Greece. \\
$^3$ \> \small
School of Physics and Astronomy, University of Birmingham, Birmingham, U.K. \\
$^4$ \> \small
CERN - European Organization for Nuclear Research, Geneva, Switzerland. \\
$^5$ \> \small
IHEP, Protvino, Russia. \\
$^6$ \> \small
IISN, Belgium. \\
$^7$ \> \small
JINR, Dubna, Russia. \\
$^8$ \> \small
High Energy Accelerator Research Organization (KEK), Tsukuba, Ibaraki 305,
Japan. \\
$^{9}$ \> \small
Faculty of Engineering, Miyazaki University, Miyazaki, Japan. \\
$^{10}$ \> \small
Oslo University, Oslo, Norway. \\
$^{11}$ \> \small
Faculty of Science, Tohoku University, Aoba-ku, Sendai 980, Japan. \\
$^{12}$ \> \small
Faculty of Science, Yamagata University, Yamagata 990, Japan. \\
\end{tabbing}
\end{titlepage}
\setcounter{page}{2}
\bigskip
\par
The study of the \pipizer system
produced in $p \overline p$
annihilations~\cite{ppbar} and
$\pi$ induced
interactions~\cite{piin,pi0pi0} has revealed that the
$\pi \pi$ S-wave has a complicated structure which indicates the
presence of several scalar states.
Some of these states may have a significant gluonic component.
In order to understand more fully the gluonic nature of these
scalar mesons a systematic study is required using as many decay modes
and production processes as possible.
This paper presents a study of the
\pipizer
final state formed in central $pp$ interactions which
are predicted to be a source of gluonic final states
via double Pomeron exchange~\cite{closerev}
and gives new information which may help in understanding the
scalar meson spectrum.
\par
The reaction
\begin{equation}
pp \rightarrow p_{f} (\pi^0 \pi^0) p_{s}
\label{eq:b}
\end{equation}
has been studied at 450~GeV/c.
The subscripts $f$ and $s$ indicate the
fastest and slowest particles in the laboratory respectively.
The WA102 experiment
has been performed using the CERN Omega Spectrometer,
the layout of which is
described in ref.~\cite{WADPT}.
Reaction (1)
has been isolated
from the sample of events having two
outgoing
charged tracks and four $\gamma$s reconstructed in the GAMS-4000
calorimeter,
by first imposing the following cuts on the components of the
missing momentum:
$|$missing~$P_{x}| <  10.0$ GeV/c,
$|$missing~$P_{y}| <  0.25$ GeV/c and
$|$missing~$P_{z}| <  0.25$ GeV/c,
where the $x$ axis is along the beam
direction.
A correlation between
pulse-height and momentum
obtained from a system of
scintillation counters was used to ensure that the slow
particle was a proton.
\par
Fig.~\ref{fi:1}a) shows the two photon mass spectrum
for $4\gamma$-events  when
the mass of the other $\gamma$-pair lies
within a band around the $\pi^0$ mass (85--185 MeV).
A clear $\pi^0$ signal is observed with a small background.
Events belonging to
reaction (\ref{eq:b}) have been selected using a
kinematical fit (6C fit, four-momentum
conservation being used and the masses of two $\pi^0$s being fixed).
By requiring that the $\chi^2$ value
for the $\pi^0\pi^0$ hypothesis is $\chi^2<12.6$,
a total of 208 452 events
have been selected.
\par
The $p_f\pi^0$ effective mass spectrum (fig.~\ref{fi:1}b))
shows a clear $\Delta^{+}(1232)$ signal which
has been removed
by requiring $M(p_f\pi^0)$~$>$~1.5~GeV.
There is no evidence for $\Delta$ production in the
$p_s\pi^0$ effective mass spectrum (not shown).
\par
The resulting centrally produced \pipizer effective mass
distribution is shown
in fig.~\ref{fi:1}c) and consists of 166 597 events.
A peak corresponding to the $f_2(1270)$ and a
sharp drop at 1~GeV can be observed.
\par
A Partial Wave Analysis (PWA) of the centrally produced \pipizer system
has been
performed assuming the \pipizer system is produced by the
collision of two particles (referred to as exchanged particles) emitted
by the scattered protons.
The $z$ axis
is defined by the momentum vector of the
exchanged particle with the greatest four-momentum transferred
in the \pipizer centre of mass.
The $y$ axis is defined
by the cross product of the two exchanged particles in the $pp$ centre of mass.
The two variables needed to specify the decay process were taken as the polar
and azimuthal angles ($\theta$, $\phi$) of the $\pi^0$ in the \pipizer
centre of mass relative to the coordinate system described above.
\par
The acceptance corrected moments $\sqrt{4\pi}t_{LM}$, defined by
\begin{equation}
I(\Omega) =  \sum_L t_{L0} Y^0_L(\Omega) +
2 \sum_{L,M >0} t_{LM}Re\{Y^M_L(\Omega)\}
\end{equation}
have been rescaled to the total number of observed events and
are shown in fig.~\ref{fi:2}. The moments
with \hbox{$M>2$} (i.e. $t_{43}$ and $t_{44}$) and all the moments
with \hbox{$L>4$} (not shown) are small and hence only
partial waves with spin \hbox{$l=0$} and 2 and absolute values of spin
$z$-projection \hbox{$m\leq1$} have been included in the PWA.
\par
An interesting feature of the moments is the presence of some structure
in the $L=2$ and $L=4$ moments for \pipizer masses below 1~GeV which
indicates the presence of $D$-waves.
This type of structure has not been observed in $\pi$ induced
reactions~\cite{piin,pi0pi0}.
In order to see if this effect is due to acceptance problems or
problems due to non-central events, we have reanalysed the data
using a series of different cuts.
Firstly, we required that $M(p_f\pi^0)$~$>$~2.0~GeV; however,
it was found that after acceptance
correction the moments were compatible with the set for
$M(p_f\pi^0)$~$>$~1.5~GeV, showing that diffractive resonances in the range
1.5~$<$~$M(p_f\pi^0)$~$<$~2.0~GeV have a negligible effect on the moments.
We have also required that the rapidity gap between any proton and
$\pi^0$ in the event is greater than 2 units. Again the resulting
acceptance corrected moments did not change.
In order to investigate any systematic effects we have also analysed the
central $\pi^+\pi^-$ data and a similar structure is also found
in the $L$~=~2 and $L$~=~4 moments~\cite{WA102pipi}.
\par
It is interesting to compare the above result with
previously published data on other centrally produced $\pi \pi$ systems.
In preliminary results from the E690 experiment
at Fermilab, activity at a similar level is observed in the $t_{40}$ moment
below 1~GeV~\cite{E690pipi}.
There is also evidence for this structure in the $\pi^0\pi^0$
data from the CERN NA12/2 experiment~\cite{NA122pi0pi0}.
The AFS experiment at the CERN
ISR also observed moments that the $t_{40}$ moment deviated from zero
in this mass region; however, in their analysis they
claimed this deviation was due to problems of the Monte Carlo simulating
low energy tracks~\cite{AFS}.
\par
This structure does indeed seem to be a real effect
which is present in centrally produced $\pi \pi$ systems.
It has recently been suggested~\cite{fec,jmf}
that central production may be due to the
fusion of two vector particles and this may explain why
higher angular momentum systems can be produced at lower masses.
\par
The amplitudes used for the PWA are defined in the reflectivity
basis~\cite{reflectivity}.
In this basis the angular distribution is given by a sum of two non-interfering
terms corresponding to negative and positive values of reflectivity.
The waves used were of the form $J^\varepsilon _m$ with $J$~$=$~$S$
and $D$,
$m$~$=$~$0,1$ and reflectivity $\varepsilon$~=~$\pm 1$.
The expressions relating the moments
($t_{LM}$) and the waves ($J^\varepsilon _m$) are given in table~\ref{ta:a}.
Since the overall phase for each reflectivity is indeterminate,
one wave in each reflectivity can be set to be real ($S_0^-$ and $D_1^+$
for example) and hence two phases can be set to zero ($\phi_{S_0^-}$ and
$\phi_{D_1^+}$ have been chosen).
This results in 6 parameters to be determined from the fit to the
angular distributions.
\par
The PWA has been performed independently in 40~MeV intervals of the \pipizer
mass spectrum. In each mass an event-by-event maximum likelihood
method has been used. The function
\begin{equation}
F=-\sum_{i=1}^Nln\{I(\Omega)\} + \sum_{L,M}t_{LM}\epsilon_{LM}
\end{equation}
has been minimised, where $N$ is the number of events in a given mass bin,
$\epsilon_{LM}$ are the efficiency corrections calculated
in the centre of the bin
and $t_{LM}$ are the moments of the angular distribution.
The moments calculated from the partial amplitudes
are shown superimposed on the experimental moments
in fig~\ref{fi:2}. As can be seen the results of the fit
well reproduce the experimental moments.
\par
The system of equations which express the moments via the partial wave
amplitudes is non-linear which leads to inherent
ambiguities. For a system with S and D waves there are two solutions
for each mass bin.
In each mass bin
one of these solutions
is found from the fit to the experimental angular distributions,
the other one can then be calculated by the method described in
ref.~\cite{reflectivity}.
In the case under study the bootstrapping  procedure is trivial
because the Barrelet
function has only two roots and their real and
imaginary parts do not cross zero as functions of mass,
as seen in fig~\ref{fi:3}a) and b).
In order to link the solutions in adjacent mass bins,
the roots are sorted over
real parts in each bin in such a way that
the real part of the first root should be
larger than the real part of the second root
(real parts of the two roots have different signs).
For the first solution the imaginary parts of both roots are
taken positive, the second solution is obtained by
complex conjugation of one of the roots.
\par
The two PWA solutions are shown in fig.~\ref{fi:4}.
By definition both solutions give identical moments and identical
values of the likelihood.
The only way to differentiate between the solutions, if different, is to
apply some external physical test, such as requiring
that at threshold the $S$-wave is the dominant wave.
For one solution
the $D$-waves are the dominant contribution
near threshold. This solution has been rejected as unphysical.
The $S$-wave for the
physical solution is characterised by a broad enhancement
below 1 GeV and two shoulders, at 1 and 1.4 GeV. Broad enhancements are
also seen in the three $D$-waves at low mass, but the intensities of
the $D$-waves near threshold are several times smaller than that
of the $S$-wave.
A peak corresponding to the $f_2(1270)$ is
clearly seen in the $D_0^-$ wave,
such a peak is less prominent in the $D_1^-$ wave
and is absent in
the $D_1^+$ wave.
\par
In order to obtain a satisfactory fit
to the $S_0^-$ wave from threshold to 2~GeV it has been found to be
necessary to use
three interfering Breit-Wigners to describe the $f_0(980)$, $f_0(1300)$
and $f_0(1500)$ and a background
of the form
$a(m-m_{th})^{b}exp(-cm-dm^{2})$, where
$m$ is the
\pipizer
mass,
$m_{th}$ is the
\pipizer
threshold mass and
a, b, c, d are fit parameters.
The Breit-Wigners have been convoluted with a Gaussian to
account for the experimental mass resolution.
The fit is shown in fig.~\ref{fi:3}c) for the entire mass range
and in fig.~\ref{fi:3}d) for masses above 1 GeV.
The parameters used to describe the
$f_0(1300)$
and $f_0(1500)$ are those found from a fit to the \pipi mass
spectrum of the same experiment~\cite{WA102pipi}, namely
\begin{tabbing}
asada \=adfsfsf99ba \=Mas \= == \=1224 \=MeVswfw, \=gaa \=  == \=1224  \=MeV
\kill
\>$f_0(1300)$ \>M \>=\>1308\>MeV,\>$\Gamma$\>=\>222\>MeV \\
\>$f_0(1500)$ \>M \>=\>1502\>MeV,\>$\Gamma$\>=\>131\>MeV.
\end{tabbing}
The parameters found for the $f_0(980)$ are
\begin{tabbing}
aaa\=adfsfsf99ba \=Mas \= == \=1224 \=pm \=1200 \=MeVswfw, \=gaa \=  == \=1224
\=pm \=1200  \=MeV   \kill
\>$f_0(980)$ \>M \>=\>986\>$\pm$\>10\>MeV,\>$\Gamma$\>=\>76\>$\pm$\>15\>MeV,
\end{tabbing}
and as can be seen, the fit well describes the \pipizer $S_0^-$-wave spectrum.
\par
In conclusion, a partial wave analysis of a
high statistics sample of centrally produced \pipizer events has been
performed.
An unambiguous physical solution has been found.
The S-wave is found to dominate the mass spectrum and is composed of
a broad enhancement at threshold, a sharp drop
at 1 GeV due to the interference between the $f_0(980)$
and the S-wave background, the $f_0(1300)$ and the $f_0(1500)$.
The D-wave shows evidence for the $f_2(1270)$ and a broad enhancement
below 1 GeV. It is interesting to note that the $f_2(1270)$ is
produced dominantly with m~=~0.
\begin{center}
{\bf Acknowledgements}
\end{center}
\par
This work is supported, in part, by grants from
the British Particle Physics and Astronomy Research Council,
the British Royal Society,
the Ministry of Education, Science, Sports and Culture of Japan
(grants no. 04044159 and 07044098), the Programme International
de Cooperation Scientifique (grant no. 576)
and
the Russian Foundation for Basic Research
(grants 96-15-96633 and 98-02-22032).

\newpage
\newpage
\begin{table}[h]
\caption{The moments of the angular distribution expressed in terms of the
partial waves.}
\label{ta:a}
\vspace{1in}
\begin{center}
\begin{tabular}{|ccl|} \hline
  & & \\
 $\sqrt{4 \pi}t_{00}$ & = & $|S_0^-|^2+|D_0^-|^2+|D^-_1|^2+|D^+_1|^2$\\
  & & \\
 $\sqrt{4 \pi}t_{20}$ & = & $\frac{\sqrt{5}}{7}(2|D_0^-|^2 + |D_1^-|^2 +
|D_1^+|^2)
 +2|S_0^-||D_0^-|\cos(\phi_{S_0^-}-\phi_{D_0^-})$\\
  & & \\
 $\sqrt{4 \pi}t_{21}$ & = &
$\frac{\sqrt{10}}{7}|D_1^-||D_0^-|\cos(\phi_{D_1^-}-\phi_{D_0^-})
 +\sqrt{2}|S_0^-||D_1^-|\cos(\phi_{S_0^-}-\phi_{D_1^-})$\\
  & & \\
 $\sqrt{4 \pi}t_{22}$ & = &
$\frac{\sqrt{15}}{7\sqrt{2}}(|D_1^-|^2-|D_1^+|^2)$\\
  & & \\
 $\sqrt{4 \pi}t_{40}$ & = &
$\frac{6}{7}|D_0^-|^2-\frac{4}{7}(|D_1^-|^2+|D_1^+|^2)$\\
  & & \\
 $\sqrt{4 \pi}t_{41}$ & = &
$\frac{2\sqrt{15}}{7}|D_0^-||D_1^-|\cos(\phi_{D_0^-}-\phi_{D_1^-})$\\
  & & \\
 $\sqrt{4 \pi}t_{42}$ & = & $\frac{\sqrt{10}}{7}(|D_1^-|^2-|D_1^+|^2)$\\
  & & \\ \hline
\end{tabular}
\end{center}
\end{table}
\clearpage
{ \large \bf Figures \rm}
\begin{figure}[h]
\caption{a) Effective $\gamma\gamma$ mass for $4\gamma$-events when
the mass of the other $\gamma$-pair lies within the $\pi^0$ mass band,
b) the $M(p_f\pi^0)$ and
c) the centrally produced \pipizer effective mass spectrum.
}
\label{fi:1}
\end{figure}
\noindent
\begin{figure}[h]
\caption{ The $\protect\sqrt{4\pi}t_{LM}$ moments from the data.
Superimposed as a solid histogram are the resulting moments calculated
from the PWA of the
\pipizer final state.
}
\label{fi:2}
\end{figure}
\noindent
\begin{figure}[h]
\caption{
a) The real and b) Imaginary parts of the roots (see text) as a
function of mass obtained from the PWA.
c) and d) The \pipizer $S_0^-$ wave with fit described in the text.
}
\label{fi:3}
\end{figure}
\noindent
\begin{figure}[h]
\caption{
The physical (solid circles) and unphysical (open circles) solutions
from the PWA of the \pipizer final state.\hfill
}
\label{fi:4}
\end{figure}
\newpage
\begin{figure}
\begin{center}
\mbox{\epsfig{figure=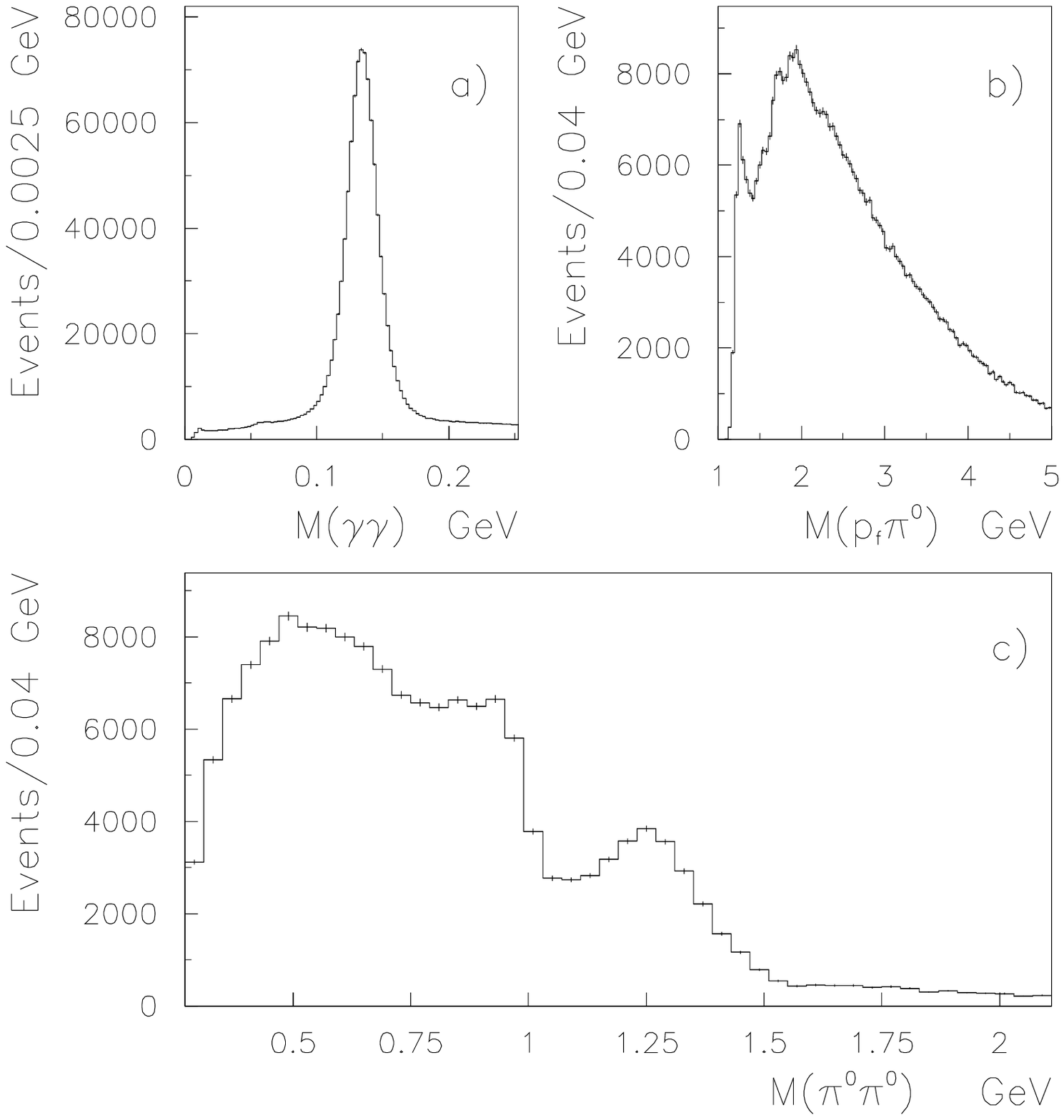,height=16cm}}
\end{center}
\begin{center} {Figure 1} \end{center}
\end{figure}
\newpage
\begin{figure}
\begin{center}
\mbox{\epsfig{figure=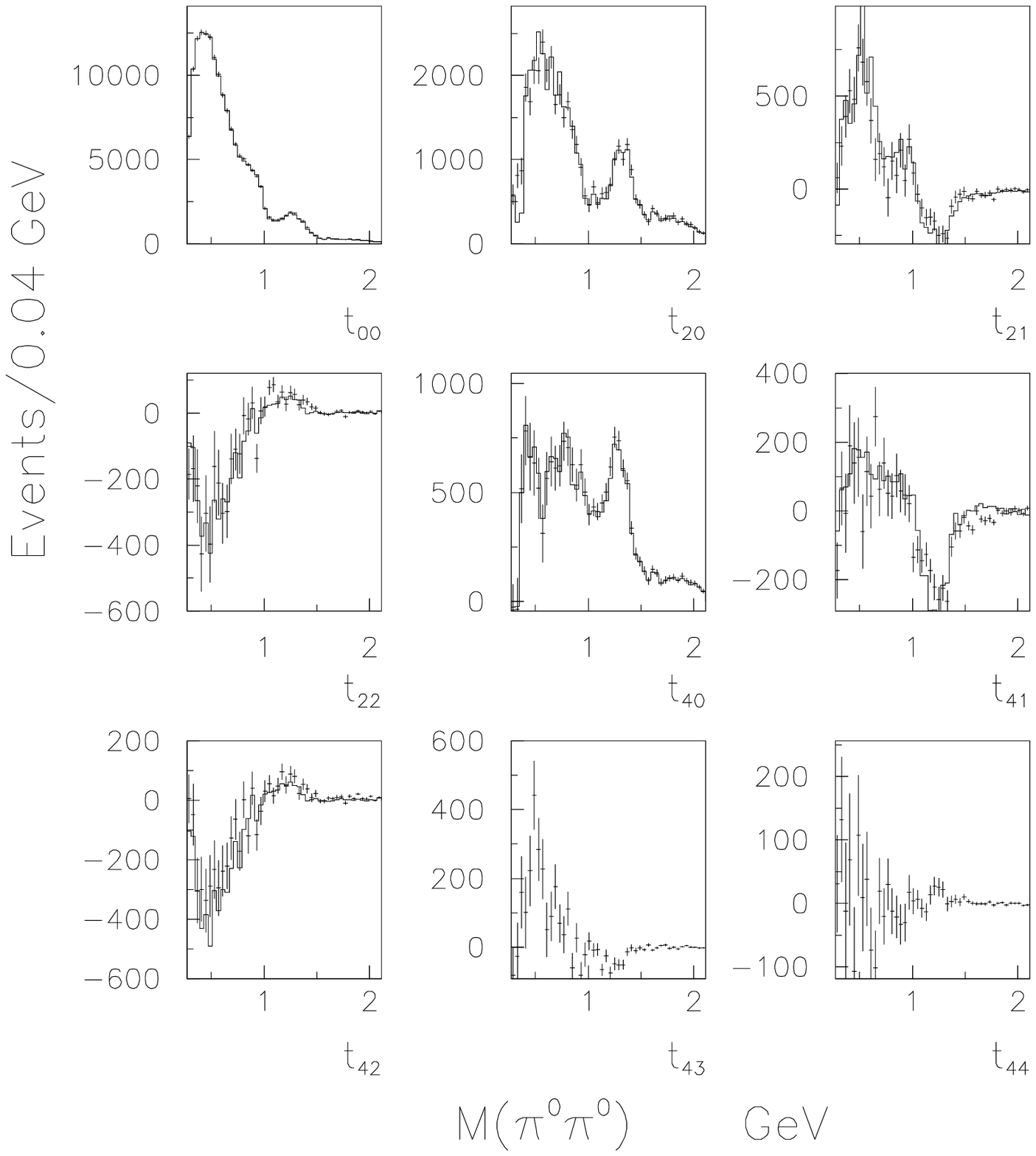,height=16cm}}
\end{center}
\begin{center} {Figure 2} \end{center}
\end{figure}
\newpage
\begin{figure}
\begin{center}
\mbox{\epsfig{figure=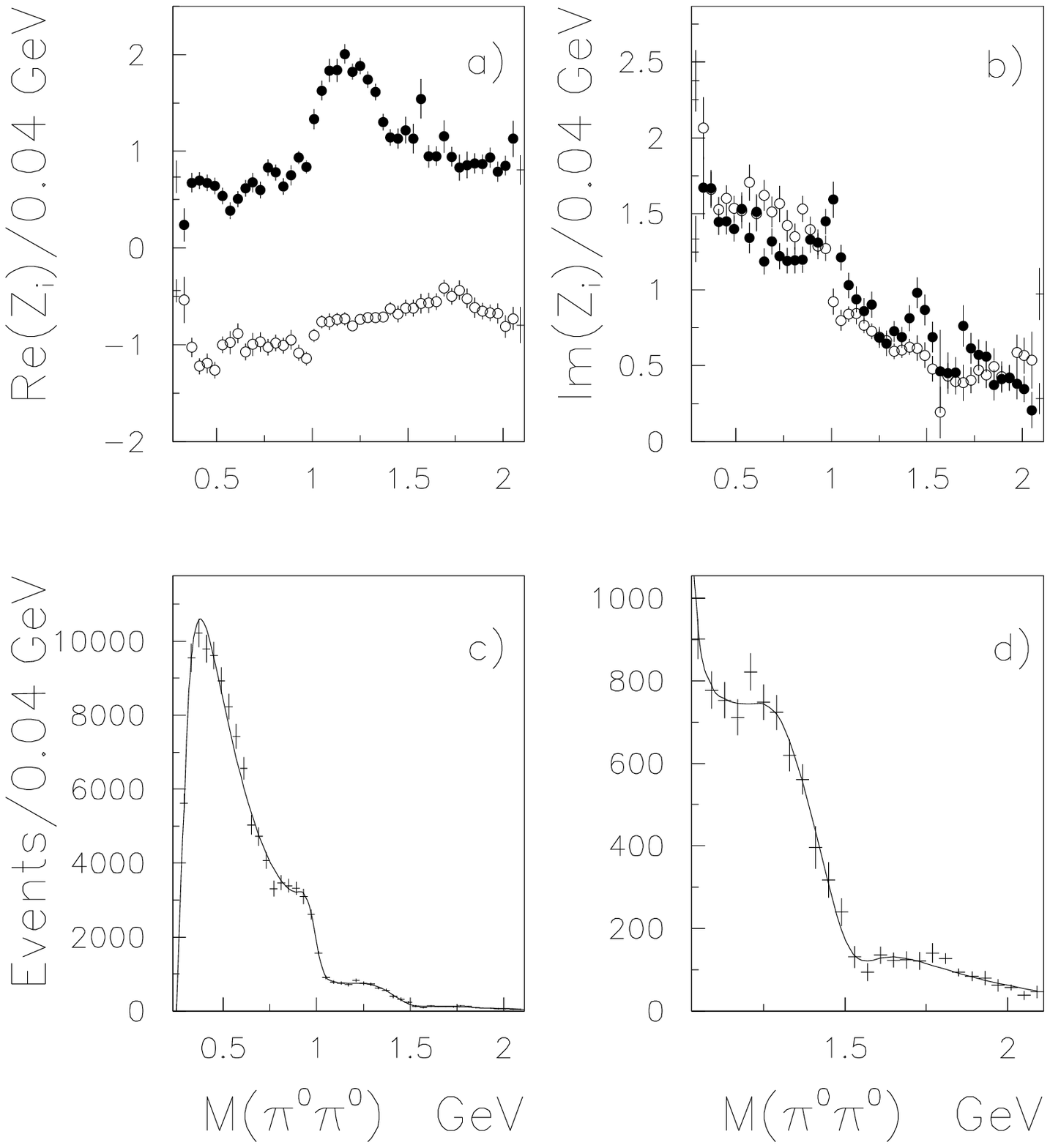,height=20cm}}
\end{center}
\begin{center} {Figure 3} \end{center}
\end{figure}
\newpage
\begin{figure}
\begin{center}
\mbox{\epsfig{figure=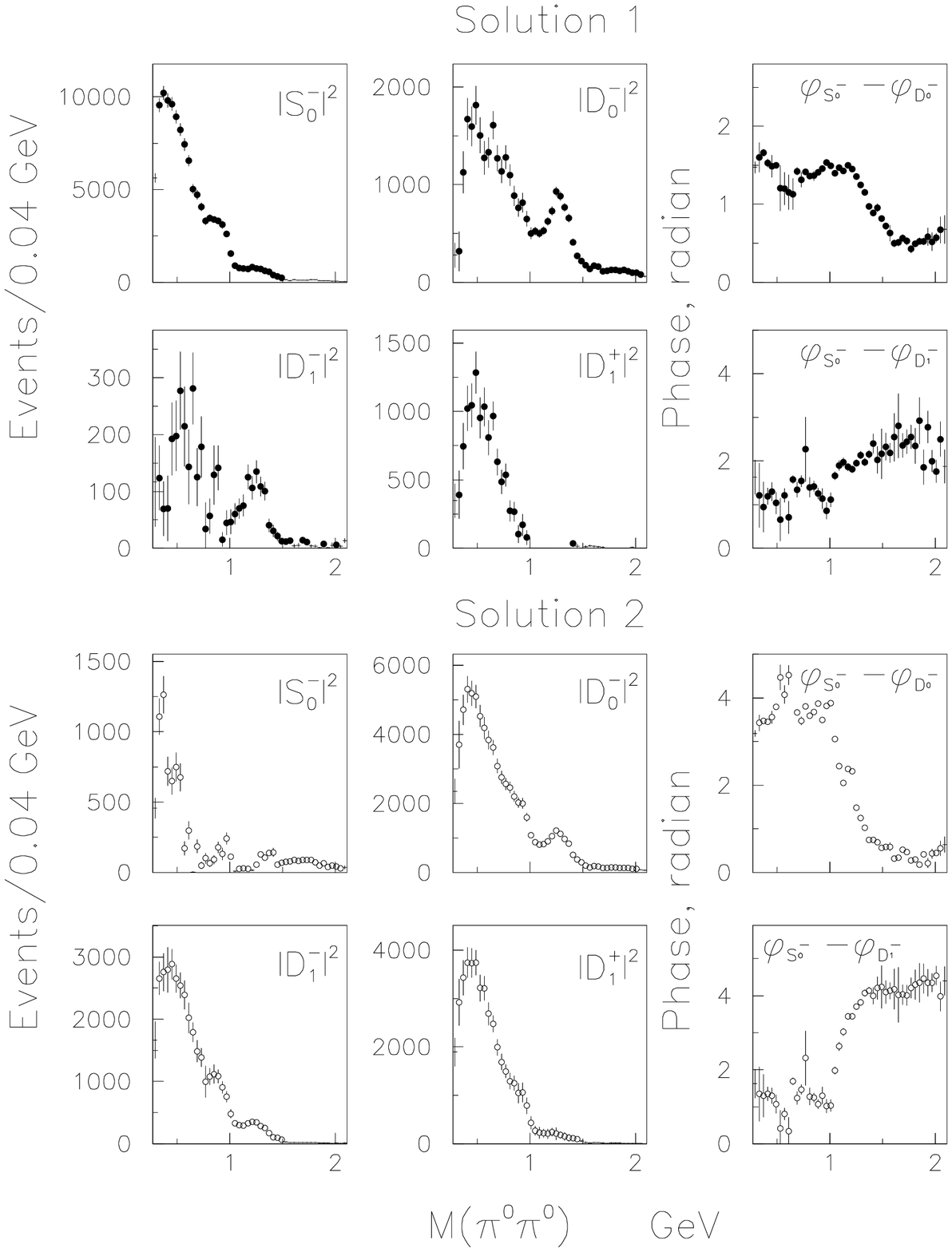,height=20cm}}
\end{center}
\begin{center} {Figure 4} \end{center}
\end{figure}
\end{document}